# Data Mining Applications: A comparative Study for Predicting Student's performance

Surjeet Kumar Yadav [1], Brijesh Bharadwaj[2], Saurabh Pal[3]
[1]Research Scholar, Shri Venkateshwara University, Moradabad
Email: surjeet_k_yadav@yahoo.co.in
[2]Assistant Professor, Dr. R. M. L. Awadh University, Faizabad India
Email: wwwbkb@rediffmail.com
[3]Head, Dept. Of MCA, VBS Purvanchal University, Jaunpur, India
Email: drsaurabhpal@yahoo.co.in

*Abstract*— **Knowledge Discovery and Data Mining (KDD) is a multidisciplinary area focusing upon methodologies for extracting useful knowledge from data and there are several useful KDD tools to extracting the knowledge. This knowledge can be used to increase the quality of education. But educational institution does not use any knowledge discovery process approach on these data. Data mining can be used for decision making in educational system. A decision tree classifier is one of the most widely used supervised learning methods used for data exploration based on divide & conquer technique. This paper discusses use of decision trees in educational data mining. Decision tree algorithms are applied on students' past performance data to generate the model and this model can be used to predict the students' performance. It helps earlier in identifying the dropouts and students who need special attention and allow the teacher to provide appropriate advising/counseling.**

I. INTRODUCTION

Students are main assets of universities/ Institutions. The students' performance plays an important role in producing the best quality graduates and post-graduates who will become great leader and manpower for the country thus responsible for the country's economic and social development. The performance of students in universities should be a concern not only to the administrators and educators, but also to corporations in the labour market. Academic achievement is one of the main factors considered by the employer in recruiting workers especially the fresh graduates. Thus, students have to place the greatest effort in their study to obtain a good grade in order to fulfil the employer's demand. Students' academic achievement is measured by the Cumulative Grade Point Average (CGPA). CGPA shows the overall students' academic performance where it considers the average of all examinations' grade for all semesters during the tenure in university. Many factors could act as barrier and catalyst to students achieving a high CGPA that reflects their overall academic performance.

The advent of information technology in various fields has lead the large volumes of data storage in various formats like students' data, teachers' data, alumni data, resource data etc. The data collected from different applications require proper method of extracting knowledge from large repositories for better decision making. Knowledge discovery in databases (KDD), often called data mining, aims at the discovery of useful information from large collections of data [1]. The main functions of data mining are applying various methods and algorithms in order to discover and extract patterns of stored data [2]. Data mining tools predict patterns, future trends and behaviors, allowing businesses to effect proactive, knowledge-driven decisions. The automated, prospective analyses offered by data mining move beyond the analysis of past events provided by retrospective tools typical of decision support systems..

There are increasing research interests in using data mining in education. This new emerging field, called Educational Data Mining, concerns with developing methods that discover knowledge from data originating from educational environments [3]. Educational Data Mining uses many techniques such as Decision Trees, Neural Networks, Naïve Bayes, K-Nearest neighbour, and many others.

The main objective of this paper is to use data mining methodologies to study students' performance in the courses. Data mining provides many tasks that could be used to study the students performance. In this research, the classification task is used to evaluate student's performance and as there are many approaches that are used for data classification, the decision tree method is used here. Student's information like Attendance, Class test, Seminar and Assignment marks were collected from the student's management system, to predict the performance at the end of the semester examination. This paper investigates the accuracy of different Decision tree.

## II. BACKGROUND AND RELATED WORKS

Data mining techniques can be used in educational field to enhance our understanding of learning process to focus on identifying, extracting and evaluating variables related to the learning process of students as described by Alaa el-Halees [4]. Mining in educational environment is called Educational Data Mining.

Han and Kamber [3] describes data mining software that allow the users to analyze data from different dimensions, categorize it and summarize the relationships which are identified during the mining process.

Bhardwaj and Pal [13] conducted study on the student performance based by selecting 300 students from 5 different degree college conducting BCA (Bachelor of Computer Application) course of Dr. R. M. L. Awadh University, Faizabad, India. By means of Bayesian classification method on 17 attributes, it was found that the factors like students' grade in senior secondary exam, living location, medium of teaching, mother's qualification, students other habit, family annual income and student's family status were highly correlated with the student academic performance.

Pandey and Pal [5] conducted study on the student performance based by selecting 600 students from different colleges of Dr. R. M. L. Awadh University, Faizabad, India. By means of Bayes Classification on category, language and background qualification, it was found that whether new comer students will performer or not.

Hijazi and Naqvi [6] conducted as study on the student performance by selecting a sample of 300 students (225 males, 75 females) from a group of colleges affiliated to Punjab university of Pakistan. The hypothesis that was stated as "Student's attitude towards attendance in class, hours spent in study on daily basis after college, students' family income, students' mother's age and mother's education are significantly related with student performance" was framed. By means of simple linear regression analysis, it was found that the factors like mother's education and student's family income were highly correlated with the student academic performance.

Khan [7] conducted a performance study on 400 students comprising 200 boys and 200 girls selected from the senior secondary school of Aligarh Muslim University, Aligarh, India with a main objective to establish the prognostic value of different measures of cognition, personality and demographic variables for success at higher secondary level in science stream. The selection was based on cluster sampling technique in which the entire population of interest was divided into groups, or clusters, and a random sample of these clusters was selected for further analyses. It was found that girls with high socio-economic status had relatively higher academic achievement in science stream and boys with low socioeconomic status had relatively higher academic achievement in general.

Z. J. Kovacic [15] presented a case study on educational data mining to identify up to what extent the enrolment data can be used to predict student's success. The algorithms CHAID and CART were applied on student enrolment data of information system students of open polytechnic of New Zealand to get two decision trees classifying successful and unsuccessful students. The accuracy obtained with CHAID and CART was 59.4 and 60.5 respectively.

Galit [8] gave a case study that use students data to analyze their learning behavior to predict the results and to warn students at risk before their final exams.

Al-Radaideh, et al [9] applied a decision tree model to predict the final grade of students who studied the C++ course in Yarmouk University, Jordan in the year 2005. Three different classification methods namely ID3, C4.5, and the NaïveBayes were used. The outcome of their results indicated that Decision Tree model had better prediction than other models.

Baradwaj and Pal [16] obtained the university students data like attendance, class test, seminar and assignment marks from the students' previous database, to predict the performance at the end of the semester.

Ayesha, Mustafa, Sattar and Khan [11] describe the use of k-means clustering algorithm to predict student's learning activities. The information generated after the implementation of data mining technique may be helpful for instructor as well as for students.

Pandey and Pal [11] conducted study on the student performance based by selecting 60 students from a degree college of Dr. R. M. L. Awadh University, Faizabad, India. By means of association rule they find the interestingness of student in opting class teaching language.

Bray [12], in his study on private tutoring and its implications, observed that the percentage of students receiving private tutoring in India was relatively higher than in Malaysia, Singapore, Japan, China and Sri Lanka. It was also observed that there was an enhancement of academic performance with the intensity of private tutoring and this variation of intensity of private tutoring depends on the collective factor namely socioeconomic conditions.

### III. DECISION TREE INTRODUCTION

A decision tree is a flow-chart-like tree structure, where each internal node is denoted by rectangles, and leaf nodes are denoted by ovals. All internal nodes have two or more child nodes. All internal nodes contain splits, which test the value of an expression of the attributes. Arcs from an internal node to its children are labelled with distinct outcomes of the test. Each leaf node has a class label associated with it.

The decision tree classifier has two phases [3]:
i) Growth phase or Build phase.
ii) Pruning phase.

The tree is built in the first phase by recursively splitting the training set based on local optimal criteria until all or most of the records belonging to each of the partitions bearing the same class label. The tree may overfit the data.

The pruning phase handles the problem of over fitting the data in the decision tree. The prune phase generalizes the tree by removing the noise and outliers. The accuracy of the classification increases in the pruning phase.

TABLE I :
FREQUENCY USAGE OF DECISION TREE ALGORITHMS

| Algorithm | Usage frequency (%) |
|---|---|
| CLS | 9 |
| ID3 | 68 |
| IDE3+ | 4.5 |
| C4.5 | 54.55 |
| C5.0 | 9 |
| CART | 40.9 |
| Random Tree | 4.5 |
| Random Forest | 9 |
| SLIQ | 27.27 |
| Public | 13.6 |
| OCI | 4.5 |
| Clouds | 4.5 |

Pruning phase accesses only the fully grown tree. The growth phase requires multiple passes over the training data. The time needed for pruning the decision tree is very less compared to build the decision tree. The table I specified represents the usage frequency of various decision tree algorithms [17]. Observing the above table the most frequently used decision tree algorithms are ID3, C4.5 and CART. Hence, the experiments are conducted on the above three algorithms.

*A. ID3 (Iterative Dichotomiser 3)*

This is a decision tree algorithm introduced in 1986 by Quinlan Ross [14]. It is based on Hunts algorithm. The tree is constructed in two phases. The two phases are tree building and pruning.

ID3 uses information gain measure to choose the splitting attribute. It only accepts categorical attributes in building a tree model. It does not give accurate result when there is noise. To remove the noise pre-processing technique has to be used.

To build decision tree, information gain is calculated for each and every attribute and select the attribute with the highest information gain to designate as a root node. Label the attribute as a root node and the possible values of the attribute are represented as arcs. Then all possible outcome instances are tested to check whether they are falling under the same class or not. If all the instances are falling under the same class, the node is represented with single class name, otherwise choose the splitting attribute to classify the instances.

Continuous attributes can be handled using the ID3 algorithm by discretizing or directly, by considering the values to find the best split point by taking a threshold on the attribute values. ID3 does not support pruning.

*B. C4.5*

This algorithm is a successor to ID3 developed by Quinlan Ross [14]. It is also based on Hunt's algorithm.C4.5 handles both categorical and continuous attributes to build a decision tree. In order to handle continuous attributes, C4.5 splits the attribute values into two partitions based on the selected threshold such that all the values above the threshold as one child and the remaining as another child. It also handles missing attribute values. C4.5 uses Gain Ratio as an attribute selection measure to build a decision tree. It removes the biasness of information gain when there are many outcome values of an attribute.

At first, calculate the gain ratio of each attribute. The root node will be the attribute whose gain ratio is maximum. C4.5 uses pessimistic pruning to remove unnecessary branches in the decision tree to improve the accuracy of classification.

*C. CART*

CART [18] stands for Classification And Regression Trees introduced by Breiman. It is also based on Hunt's algorithm. CART handles both categorical and continuous attributes to build a decision tree. It handles missing values.

CART uses Gini Index as an attribute selection measure to build a decision tree .Unlike ID3 and C4.5 algorithms, CART produces binary splits. Hence, it produces binary trees. Gini Index measure does not use probabilistic assumptions like ID3, C4.5. CART uses cost complexity pruning to remove the unreliable branches from the decision tree to improve the accuracy.

## IV. DATA MINING PROCESS

In present day's educational system, a student's performance is determined by the internal assessment and end semester examination. The internal assessment is carried out by the teacher based upon student's performance in educational activities such as class test, seminar, assignments, general proficiency, attendance and lab work. The end semester examination is one that is scored by the student in semester examination. Each student has to get minimum marks to pass a semester in internal as well as end semester examination.

*A. Data Preparations*

The data set used in this study was obtained from VBS Purvanchal University, Jaunpur (Uttar Pradesh), India on the sampling method of computer Applications department of course MCA (Master of Computer Applications) from session 2008 to 2011. Initially size of the data is 48. In this step data stored in different tables was joined in a single table after joining process errors were removed.

*B. Data Selection and Transformation*

In this step only those fields were selected which were required for data mining. A few derived variables were selected. While some of the information for the variables was extracted from the database. All the predictor and response variables which were derived from the database are given in Table II for reference.

The domain values for some of the variables were defined for the present investigation as follows:

**PSM –** Previous Semester Marks/Grade obtained in MCA course. It is split into five class values: *First – ≥60%, Second – ≥45% and < 60%, Third – ≥ 36% and < 45%, Fail < 36%.*

TABLE II :
STUDENTS RELATED VARIABLES

| Variable | Description | Possible Values |
|---|---|---|
| PSM | Previous Semester Marks | {First ≥ 60% Second ≥ 45 & <60% Third ≥ 36 & <45%, Fail < 36%} |
| CTG | Class Test Grade | {Poor , Average, Good} |
| SEM | Seminar Performance | {Poor , Average, Good} |
| ASS | Assignment | {Yes, No} |
| ATT | Attendance | {Poor , Average, Good} |
| LW | Lab Work | {Yes, No} |
| ESM | End Semester Marks | {First ≥ 60% Second ≥ 45 & <60% Third ≥ 36 & <45% Fail < 36%} |

- **CTG –** Class test grade obtained. Here in each semester two class tests are conducted and average of two class test are used to calculate sessional marks. CTG is split into three classes: *Poor – < 40%, Average – ≥ 40% and < 60%, Good –≥60%.*

- **SEM –** Seminar Performance obtained. In each semester seminar are organized to check the performance of students. Seminar performance is evaluated into three classes: *Poor – Presentation and communication skill is low, Average – Either presentation is fine or Communication skill is fine, Good – Both presentation and Communication skill is fine.*

- **ASS –** Assignment performance. In each semester two assignments are given to students by each teacher. Assignment performance is divided into two classes: *Yes – student submitted assignment, No – Student not submitted assignment.*

- **ATT –** Attendance of Student. Minimum 70% attendance is compulsory to participate in End Semester Examination. But even though in special cases low attendance students also participate in End Semester Examination on genuine reason basis. Attendance is divided into three classes: *Poor - <60%, Average - ≥ 60% and < 80%, Good - ≥ 80%.*

- **LW –** Lab Work. Lab work is divided into two classes: *Yes – student completed lab work, No – student not completed lab work*.
- **ESM -** End semester Marks obtained in MCA semester and it is declared as response variable. It is split into five class values: First – ≥ 60%, Second – ≥ 45% and <60%, Third – ≥ 36% and < 45%, Fail < 36%.

### C. Data Set

The data set of 48 students used in this study was obtained from VBS Purvanchal University, Jaunpur (Uttar Pradesh) Computer Applications department of course MCA (Master of Computer Applications) from session 2008 to 2011.

TABLE III : DATA SET

| S. No. | PSM | CTG | SEM | ASSS | ATT | LW | ESM |
|---|---|---|---|---|---|---|---|
| 1. | First | Good | Good | Yes | Good | Yes | First |
| 2. | First | Good | Average | Yes | Good | Yes | First |
| 3. | First | Good | Average | No | Average | No | First |
| 4. | First | Average | Good | No | Good | Yes | First |
| 5. | First | Average | Average | No | Good | Yes | First |
| 6. | First | Poor | Average | No | Average | Yes | First |
| 7. | First | Poor | Average | No | Poor | Yes | Second |
| 8. | First | Average | Poor | Yes | Average | No | First |
| 9. | First | Poor | Poor | No | Poor | No | Third |
| 10. | First | Average | Average | Yes | Good | No | First |
| 11. | Second | Good | Good | Yes | Good | Yes | First |
| 12. | Second | Good | Average | Yes | Good | Yes | First |
| 13. | Second | Good | Average | Yes | Good | No | First |
| 14. | Second | Average | Good | Yes | Good | No | First |
| 15. | Second | Good | Average | Yes | Average | Yes | First |
| 16. | Second | Good | Average | Yes | Poor | Yes | Second |
| 17. | Second | Average | Average | Yes | Good | Yes | Second |
| 18. | Second | Average | Average | Yes | Poor | Yes | Second |
| 19. | Second | Poor | Average | No | Good | Yes | Second |
| 20. | Second | Average | Poor | Yes | Average | Yes | Second |
| 21. | Second | Poor | Average | No | Poor | No | Third |
| 22. | Second | Poor | Poor | Yes | Average | Yes | Third |
| 23. | Second | Poor | Poor | No | Average | Yes | Third |
| 24. | Second | Poor | Poor | Yes | Good | Yes | Second |
| 25. | Second | Poor | Poor | Yes | Poor | Yes | Third |
| 26. | Second | Poor | Poor | No | Poor | Yes | Fail |
| 27. | Third | Good | Good | Yes | Good | Yes | First |
| 28. | Third | Average | Good | Yes | Good | Yes | Second |
| 29. | Third | Good | Average | Yes | Good | Yes | Second |
| 30. | Third | Good | Good | Yes | Average | Yes | Second |
| 31. | Third | Good | Good | No | Good | Yes | Second |
| 32. | Third | Average | Average | Yes | Good | Yes | Second |
| 33. | Third | Average | Average | No | Average | Yes | Third |
| 34. | Third | Average | Good | No | Good | Yes | Third |
| 35. | Third | Good | Average | No | Average | Yes | Third |
| 36. | Third | Average | Poor | No | Average | Yes | Third |
| 37. | Third | Poor | Average | Yes | Average | Yes | Third |
| 38. | Third | Poor | Average | No | Poor | Yes | Fail |
| 39. | Third | Average | Average | No | Poor | Yes | Third |
| 40. | Third | Poor | Poor | No | Good | No | Third |
| 41. | Third | Poor | Poor | No | Poor | Yes | Fail |
| 42. | Third | Poor | Poor | No | Poor | No | Fail |
| 43. | Fail | Good | Good | Yes | Good | Yes | Second |
| 44. | Fail | Good | Good | Yes | Average | Yes | Second |
| 45. | Fail | Average | Good | Yes | Average | Yes | Third |
| 46. | Fail | Poor | Poor | Yes | Average | No | Fail |
| 47. | Fail | Good | Poor | No | Poor | Yes | Fail |
| 48. | Fail | Poor | Poor | No | Poor | Yes | Fail |

### D. Model Construction

The Weka Knowledge Explorer is an easy to use graphical user interface that harnesses the power of the Weka software. The major Weka packages are Filters, Classifiers, Clusters, Associations, and Attribute Selection is represented in the Explorer along with a Visualization tool, which allows datasets and the predictions of Classifiers and Clusters to be visualized in two dimensions. The workbench contains a collection of visualization tools and algorithms for data analysis and predictive modelling together with graphical user interfaces for easy access to this functionality. It was primarily designed as a tool for analysing data from agricultural domains. Now it is used in many different application areas, in particular for educational purposes and research. The main strengths is freely available under the GNU General Public License, very portable because it is fully implemented in the Java programming language and runs on any modern computing platform, contains a comprehensive collection of data pre-processing and modelling techniques. Weka supports several standard data mining tasks like data clustering, classification, regression, pre-processing, visualization and feature selection. These techniques are predicated on the assumption that the data is available as a single flat file or relation. Each data point is described by a fixed number of attributes and an important area is currently not covered by the algorithms included in the Weka distribution is sequence modelling.

From the above data, mca.arff file was created. This file was loaded into WEKA explorer. The classify panel enables the user to apply classification and regression algorithms to the resulting dataset, to estimate the accuracy of the resulting predictive model, and to visualize erroneous predictions, or the model itself. There are 16 decision tree algorithms like ID3, J48, Simple CART etc. implemented in WEKA. The algorithm used for classification is ID3, C4.5 and CART. Under the "Test options", the 10-fold cross-validation is selected as our evaluation approach. Since there is no separate evaluation data set, this is necessary to get a reasonable idea of accuracy of the generated model. The model is generated in the form of decision tree.

*E. Results Obtained*

The Table IV shows the accuracy of ID3, C4.5 and CART algorithms for classification applied on the above data sets using 10-fold cross validation is observed as follows:

TABLE IV:
CLASSIFIERS ACCURACY

| Algorithm | Correctly Classified Instances | Incorrectly Classified Instances |
|---|---|---|
| ID3 | 52.0833% | 35.4167% |
| C4.5 | 45.8333% | 54.1667 % |
| CART | 56.25% | 43.75% |

Table IV shows that a CART technique has highest accuracy of 56.25% compared to other methods. ID3 algorithm also showed an acceptable level of accuracy.

The Table V shows the time complexity in seconds of various classifiers to build the model for training data.

TABLE V:
EXECUTION TIME TO BUILD THE MODEL

| Algorithm | Execution Time (Sec) |
|---|---|
| ID3 | 0 |
| C4.5 | 0.02 |
| CART | 0.05 |

The classification matrix has been presented in Table VI, VII and VIII, which compared the actual and predicted classifications. In addition, the classification accuracy for the four-class outcome categories was presented.

TABLE VI:
CLASSIFICATION MATRIX-ID3 PREDICTION MODEL

| ESM | | Predicted | | | | % of correct Precision |
|---|---|---|---|---|---|---|
| | | First | Second | Third | Fail | |
| Actual | First | 8 | 3 | 0 | 0 | 66.7% |
| | Second | 4 | 6 | 2 | 0 | 42.9% |
| | Third | 0 | 4 | 7 | 2 | 70.00% |
| | Fail | 0 | 1 | 1 | 4 | 66.7% |

TABLE VII:
CLASSIFICATION MATRIX-C4.5 PREDICTION MODEL

| ESM | | Predicted | | | | % of correct Precision |
|---|---|---|---|---|---|---|
| | | First | Second | Third | Fail | |
| Actual | First | 8 | 4 | 2 | 0 | 55.31% |
| | Second | 3 | 8 | 2 | 1 | 47.1% |
| | Third | 4 | 4 | 4 | 1 | 30.8% |
| | Fail | 0 | 1 | 5 | 1 | 33.3% |

TABLE VIII:
CLASSIFICATION MATRIX-CART PREDICTION MODEL

| ESM | | Predicted | | | | % of correct Precision |
|---|---|---|---|---|---|---|
| | | First | Second | Third | Fail | |
| Actual | First | 9 | 3 | 2 | 0 | 69.2% |
| | Second | 2 | 10 | 2 | 0 | 55.6% |
| | Third | 2 | 4 | 5 | 2 | 41.7% |
| | Fail | 0 | 1 | 3 | 3 | 60.0% |

The knowledge represented by decision tree can be extracted and represented in the form of IF-THEN rules.

| IF PSM = 'First' AND ATT = 'Good' AND CTG = 'Good' or 'Average' THEN ESM = First |
|---|
| IF PSM = 'First' AND CTG = 'Good' AND ATT = "Good' OR 'Average' THEN ESM = 'First' |
| IF PSM = 'Second' AND ATT = 'Good' AND ASS = 'Yes' THEN ESM = 'First' |
| IF PSM = 'Second' AND CTG = 'Average' AND LW = 'Yes' THEN ESM = 'Second' |
| IF PSM = 'Third' AND CTG = 'Good' OR 'Average' AND ATT = "Good' OR 'Average' THEN PSM = 'Second' |
| IF PSM = 'Third' AND ASS = 'No' AND ATT = 'Average' THEN PSM = 'Third' |
| IF PSM = 'Fail' AND CTG = 'Poor' AND ATT = 'Poor' THEN PSM = 'Fail' |

Fig 1. Rule Set generated by Decision Tree

The classifiers accuracy on various data sets is represented in the form of a graph.

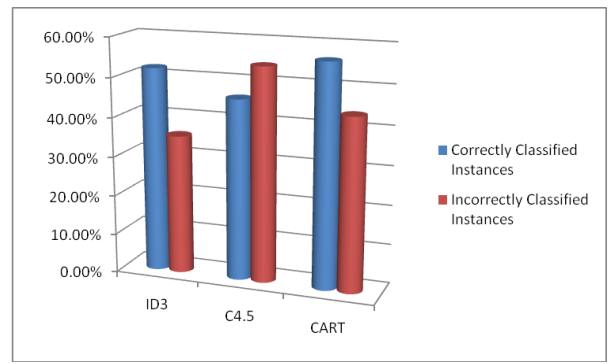

Fig 2. Comparison of Classifiers

V. CONCLUSION

Data Mining is gaining its popularity in almost all applications of real world. One of the data mining techniques i.e., classification is an interesting topic to the researchers as it is accurately and efficiently classifies the data for knowledge discovery. Decision trees are so popular because they produce classification rules that are easy to interpret than other classification methods. Frequently used decision tree classifiers are studied and the experiments are conducted to find the best classifier for Student data to predict the student's performance in the end semester examination. The experimental results show that CART is the best algorithm for classification of data.

This study will help to the students and the teachers to improve the performance of the students. This study will also work to identify those students which needed special attention and will also work to reduce fail ratio and taking appropriate action for the next semester examination.